\documentclass{PoS}

\usepackage{graphicx} 
\usepackage{epsfig}
\usepackage{microtype}

\title{The Milky Way Heart: Investigating molecular gas and gamma-ray morphologies in the Central Molecular Zone}

\ShortTitle{Molecular gas and gamma-ray morphologies in the Central Molecular Zone}

\author{David I. Jones$^{a}$, Michael Burton$^b$,  Paul Jones$^{b,c}$, Andrew Walsh$^d$, Gavin Rowell$^e$, and Felix Aharonian$^{f,a}$ \\
\llap{$^a$}Max Planck Institut f\"{u}r Kernphysik, Postfach 103980, 69029 Heidelberg, Germany\\
\llap{$^b$}School of Physics, University of New South Wales, 2052, Sydney, Australia.\\
\llap{$^c$}Dept. de Astronom\'{i}a, Universidad de Chile, Casilla 36-D, Santiago, Chile. \\
\llap{$^d$}Centre for Astronomy, James Cook University, Townsville, Queensland, Australia.\\
\llap{$^e$}School of Chemistry and Physics, University of Adelaide, Adelaide, Australia, 5005. \\
\llap{$^f$}Centre for Cosmic Physics, Dublin Institute for Advanced Study, Dublin D4, Ireland. \\		
}

\abstract{

Since the discovery of a broad distribution of very high energy (VHE; $ >0.1$~TeV) gamma-rays in the Central Molecular Zone (CMZ) of the Galaxy in 2006 by the HESS collaboration, the correlation of this emission with the integrated intensity of the CS(1-0) molecular line emission has inferred a hadronic origin for the gamma-rays.  Here we describe the beginning of our investigation into the strength of this correlation utilising new multi-line millimeter data from the Mopra CMZ and HOP surveys and multi-wavelength GBT radio continuum observations towards the CMZ and compare these in detail with the diffuse TeV gamma-ray emission from HESS.  The benefit of these new data is that they allow us to simultaneously observe and analyse correlations using a large number ($>10$) of molecular species, some of which contain their isotopologue pairs.  The use of isotopologue pairs is especially powerful, since it allows one to analyse the optical depth of a number of different molecular species, thus investigating the nature of the correlation over a range of different physical conditions. Here we begin by comparing the integrated line emission and continuum radio emission with the diffuse gamma-ray emission, and, by using isotopologue pairs such as HCN/H$^{13}$CN, obtain optical depths throughout the CMZ corresponding to regions of both strong and weak gamma-ray emission. We find that the radio continuum better matches the peak of the gamma-ray emission, which corresponds to the more compact -- compared to the relatively coarse resolution of the gamma-ray images -- sources in the CMZ. 
Using the isotopologue pairs, we find that the optical depth at all positions and velocities within the CMZ are about $\tau\sim2-4$.
This is similar to that found for the CS(1--0) line and would underestimate the mass of the CMZ, potentially explaining why molecular line emission peaks appear offset from the gamma-ray peaks.
}

\FullConference{25th Texas Symposium on Relativistic Astrophysics \\
		 December 6-10, 2010\\
		 Heidelberg, Germany}

\begin{document}

\section{Introduction}
In 2004 the High Energy Stereoscopic System (HESS) gamma-ray telescope discovered several point-sources of very high energy (VHE; $>0.1$~TeV) gamma-ray emission near the dynamical centre of the Galaxy (the Galactic centre; GC \cite{Aharonian2006}).  These sources have since been shown to be coincident with the GC (either the supermassive black hole, Sgr~A* or a pulsar wind nebula \cite{Aharonian2010}), and the supernova remnant G0.9+0.1. It was then revealed -- after the subtraction of these two sources -- that there also existed a band of diffuse TeV gamma-rays pervading the central regions of our Galaxy \cite{Aharonian2006}. Furthermore, \cite{Aharonian2006} argued that this diffuse GC TeV gamma-ray emission was ``spatially correlated'' with the integrated emission of the molecular gas, as traced by the CS(1--0) emission line \cite{Tsuboi1999} out to a Galactic longitude of $\sim1^\circ$, where the correlation with the molecular material became worse (c.f. Figure 1b from \cite{Aharonian2006}). The presence of this correlation could indicate a hadronic origin for the diffuse TeV gamma-ray glow at the GC, whereby the gamma-rays are produced by the interaction of CR protons and nucleons with the ambient nucleonic matter.  In this manner, TeV gamma-rays are created from the decay of neutral pions created in the collisions via $\pi^0\rightarrow\gamma\gamma$ \cite{Crocker2007}\footnote{It should be noted, however, that the correlation does not rule out a ``leptonic scenario'' -- whereby the TeV gamma-rays are created by the up-scattering of lower-energy photons from ambient light fields by in situ accelerated electrons -- as an explanation for the observed emission \cite{Aharonian2005}.}. In all of the resulting studies of the diffuse TeV emission, the inferred results rest -- crucially -- on the assumption of the strength of the correlation between the molecular gas and the diffuse TeV gamma-rays. We begin to rectify the deficit in the observational domain, where there has not been a commensurate effort to investigate the strength of the correlation between the molecular gas of the GC and (both) the (point-like and) diffuse TeV gamma-ray emission.  In this proceeding, we explore the nature of the gas in this region and how implications from the assumption of a hadronic origin to the diffuse gamma-rays are manifest in the molecular-line and TeV gamma-ray data.

\begin{figure}[h]
   \centering
\epsfig{file=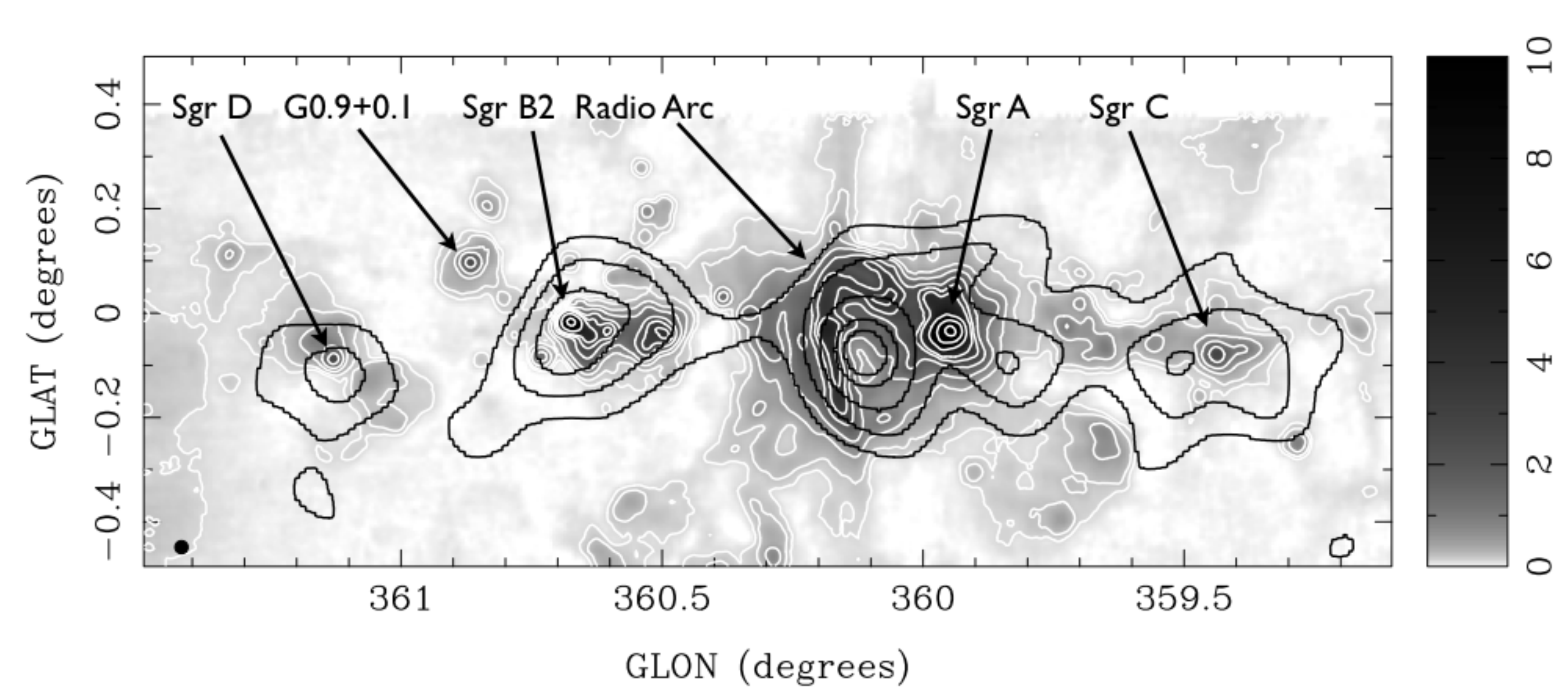,width=\columnwidth}
   \caption{The radio continuum 8.5 GHz view of the Galactic centre \cite{Law2008} with the major sources labelled and units of Jy/beam and associated (white) contours.  The black HESS contours at 290, 300, 320, 340 and 360 excess counts. The resolution of the radio continuum is $3.5'$, and is indicated by the beam in the lower, left-hand corner.}
\label{fig:radio}
\end{figure}

\subsection{The data}
\subsubsection{The radio continuum data}
The radio continuum study of the Galactic centre by \cite{Law2008} was designed for spectral index studies of the region roughly corresponding to the CMZ.  They used the Greenbank radio telescope (GBT) at 0.33, 1.4, 4.8 and 8.5~GHz to map the region with a resolution, at 8.5~GHz of $1.5'$ and an RMS sensitivity of 9~mJy/beam (see Table~1 of \cite{Law2008} for information on the survey at other wavelengths). In this work, we use only the 8.5~GHz data (see Jones D. et~al forthcoming for a more complete discussion of this data in regards to the diffuse TeV gamma-rays from the GC).

\subsubsection{The Mopra CMZ survey data}
The Mopra CMZ survey (\cite{Jones2008}, Jones P. et~al., 2011, in prep.) is a molecular line mapping survey of the Central Molecular Zone (CMZ) of the Galaxy extending approximately $3^\circ$ along the Galactic plane and $0.25^\circ$ out of the plane -- an extent of $\sim450\times150$~pc. The survey samples the CMZ at frequencies between 86 GHz and 115~GHz with the 22-m Mopra radio telescope\footnote{The Mopra radio telescope is part of the Australia Telescope which is funded by the Commonwealth of Australia for operations as a National Facility managed by CSIRO.} using the UNSW-MOPS spectrometer on Mopra, combined with on-the-fly mapping and the MMIC receiver. The survey consists of data-cubes with a velocity resolution of $\sim1$~km/s over the 8~GHz bandwidth (corresponding to a velocity range of roughly $-300$ to $+300$~km/s), an angular resolution of $\sim40''$ and an RMS noise of $T_{MB}\sim0.05$~K per channel \cite{Jones2008}.

\subsubsection{The HOPS data}
The H$_2$O Plane Survey (HOPS; \cite{Walsh2008}, Walsh et~al., 2011, in prep.) is also a molecular line mapping survey which intends to map a large portion of the southern Galaxy with the 12~mm receiver on the Mopra radio telescope. The main impetus for this survey is to observe the NH$_3$ inversion transitions, as well as the H$_2$O($6_{16}$--$5_{23}$) maser line at 22.235~GHz to locate and characterise the dense molecular cores along the (southern) Galactic plane. The survey observes frequencies between 19.5~GHz and 27.5~GHz with the 22-m Mopra radio telescope and consists of data-cubes with a velocity resolution of $\sim0.4$~km/s over the 8~GHz bandwidth (corresponding to a velocity range of roughly $-300$ to $+300$~km/s), a resolution of $\sim150''$ and an RMS noise of $T_{MB}\sim0.4$~Jy/beam \cite{Walsh2008}.

\subsubsection{The HESS data}
The High Energy Stereoscopic System (HESS) of very high energy gamma-ray telescopes is situated in Namibia, Southern Africa, and uses the atmospheric Cherenkov technique to image gamma-ray sources.  The ridge of VHE gamma-rays which we discuss here was reported in \cite{Aharonian2006} and obtained by the subtraction of the two point sources and is an excess counts image (as opposed to a significance image), which represents significant excess counts above the background of cosmic-rays and a resolution (FWHM) of $11.3'$.

\section{Morphological correspondence of different wavebands with TeV gamma-rays}

\subsection{Correspondence with radio data}
Figure \ref{fig:radio} shows the GC region at 8.5~GHz from \cite{Law2008} overlaid with (white) 8.5~GHz and (black) TeV gamma-ray contours.  This shows that, because there is no issue with optical depth of the thermal (mainly from H{\sc ii} regions) and non-thermal (mainly from supernova remnants -- SNRs -- and the Radio Arc) photons, the dense cores of molecular clouds such as Sgr~B2 correspond well with the peaks of the gamma-ray emission. However, the correspondence seems to fall down because the emission in the radio is dominated more by compact sources (for the coarse HESS beam), such as H{\sc ii} regions and SNRs, rather than a broad distribution of matter that the molecular lines trace (c.f. Figures \ref{fig:HOPS} and \ref{fig:CMZ}). We also note some correspondence between the Sgr~D SNR and a local gamma-ray peak. However, given the fact that the Sgr D SNR is likely a background object compared to the CMZ \cite{Mehringer1998}, we suggest that the gamma-ray emission from this region is not physically connected to the CMZ. 


\begin{figure}[h]
   \centering
\epsfig{file=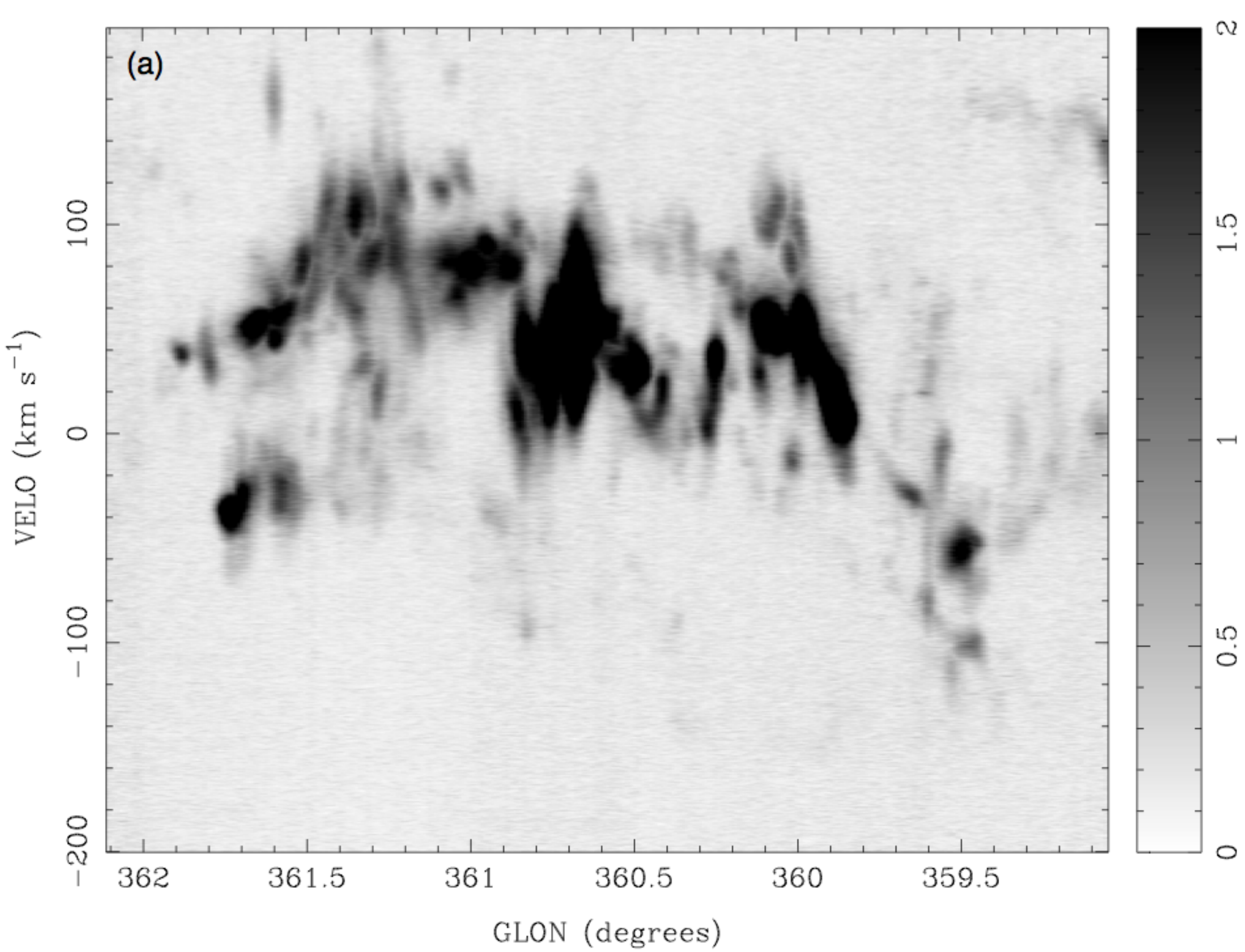,width=0.5\columnwidth}\epsfig{file=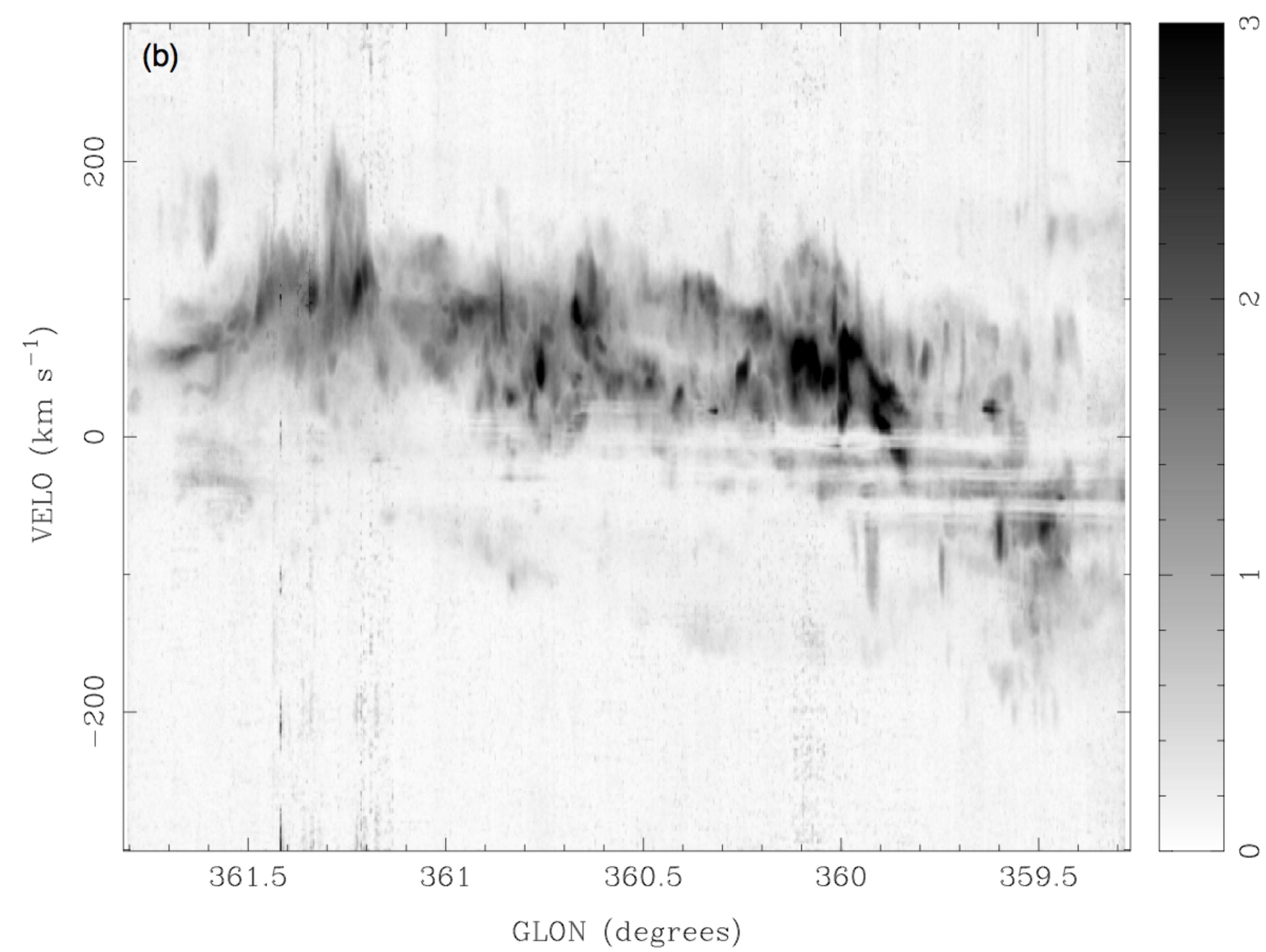,width=0.5\columnwidth}
   \caption{A velocity longitude plot of (a) NH$_3$ (3,3) rotational line and (b) HCN(1--0) emission in the GC region. Note that the resolution in the NH$_3$ images is $\sim150''$, whereas the resolution of the HCN(1--0) image is $\sim40''$, so that the distributions of the molecules are quite similar in both cases. The intensity scale on the right is temperature in K.}
   \label{fig:lvDiagrams}
\end{figure}

\subsection{Correspondence with HOPS data}
Figure \ref{fig:lvDiagrams}(a) shows a longitude--velocity ($l-v$) plot for NH$_3$ (3,3) illustrates how the NH$_3$ emission does not suffer from absorption due to local clouds, unlike the HCN(1--0) emission in Figure \ref{fig:lvDiagrams}(b) (in which the absorption is seen as horizontal streaks through velocities around 0~km/s). This does not imply, however, that the NH$_3$ data possesses the advantage that when integrating the emission along the line of sight because of the lack of absorption by foreground clouds (although, this was an important consideration in the use of the CS(1--0) emission line data from \cite{Tsuboi1999} in the original HESS publication \cite{Aharonian2006}). The NH$_3$ emission does suffer from optical depth effects, which may not be clear from Figure \ref{fig:lvDiagrams}(a). The determination of how much absorption the NH$_3$ emission suffers is complex, since the hyperfine lines are blended near the GC due to the wide line widths (as evidenced in Figure \ref{fig:lvDiagrams}).  Thus radiative transfer modelling is required for the full interpretation of the NH$_3$ data, whereas the optical depth analysis performed using the HCN(1--0) tracer and its isotopologue actually corrects for the foreground absorption of HCN.

\begin{figure}[h]
   \centering
\epsfig{file=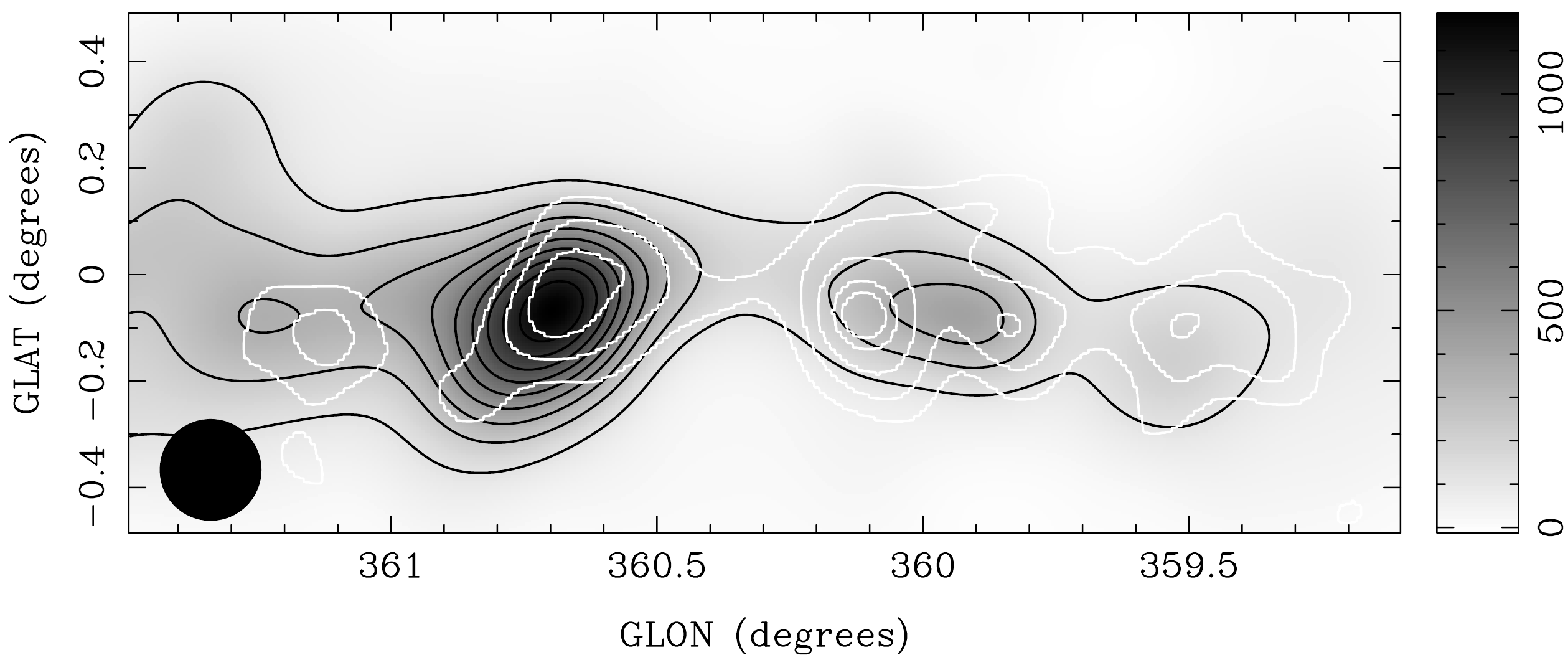,width=\columnwidth}
   \caption{Integrated emission image (over the range -200 to 200 km s$^{-1}$ in units of K km s$^{-1}$) of the NH$_3$ (2,2) rotational line. Overlaid on the total intensity are (white) HESS contours at 290, 300, 320, 340 and 360 excess counts and (black) NH$_3$(2--2) contours at 20\% to 90\% in steps of 10\%. The $11.3'$ beam, which the line emission has been convolved with is shown in the lower left-hand corner.}
   \label{fig:HOPS}
\end{figure}

Figure \ref{fig:HOPS} shows the result of integrating the NH$_3$ (2,2) inversion transition line from -200 to 200~km~s$^{-1}$ and convolving the image with a Gaussian beam of a resolution equal to that of the HESS data.  This shows that the NH$_3$ emission at temperatures and densities which this line traces (viz. $n_H\sim10^4$~cm$^{-3}$ and $T_{ex}\sim70$~K) are dominated by the Sgr~B cloud complex, although there are also peaks which are at or near the Sgr~C (at negative Galactic longitude) H{\sc ii} region and the $l=1.3^\circ$ cloud complex.

We note here that although the $l=1.3^\circ$ complex is considered part of the CMZ, the fact that it does not seem to be illuminated by TeV gamma-rays is not necessarily because of (a lack of) diffusive particles from a central source.  As alluded to above, the structure of the components of GC sources suggests that the SNR/H{\sc ii} region Sgr~D and the $l=1.3^\circ$ are on opposite sides of the GC \cite{Mehringer1998} so that the distances that CRs would have to traverse from source to target may be even larger than suspected.

\subsection{Correspondence with CMZ data}
\begin{figure}[h]
   \centering
\epsfig{file=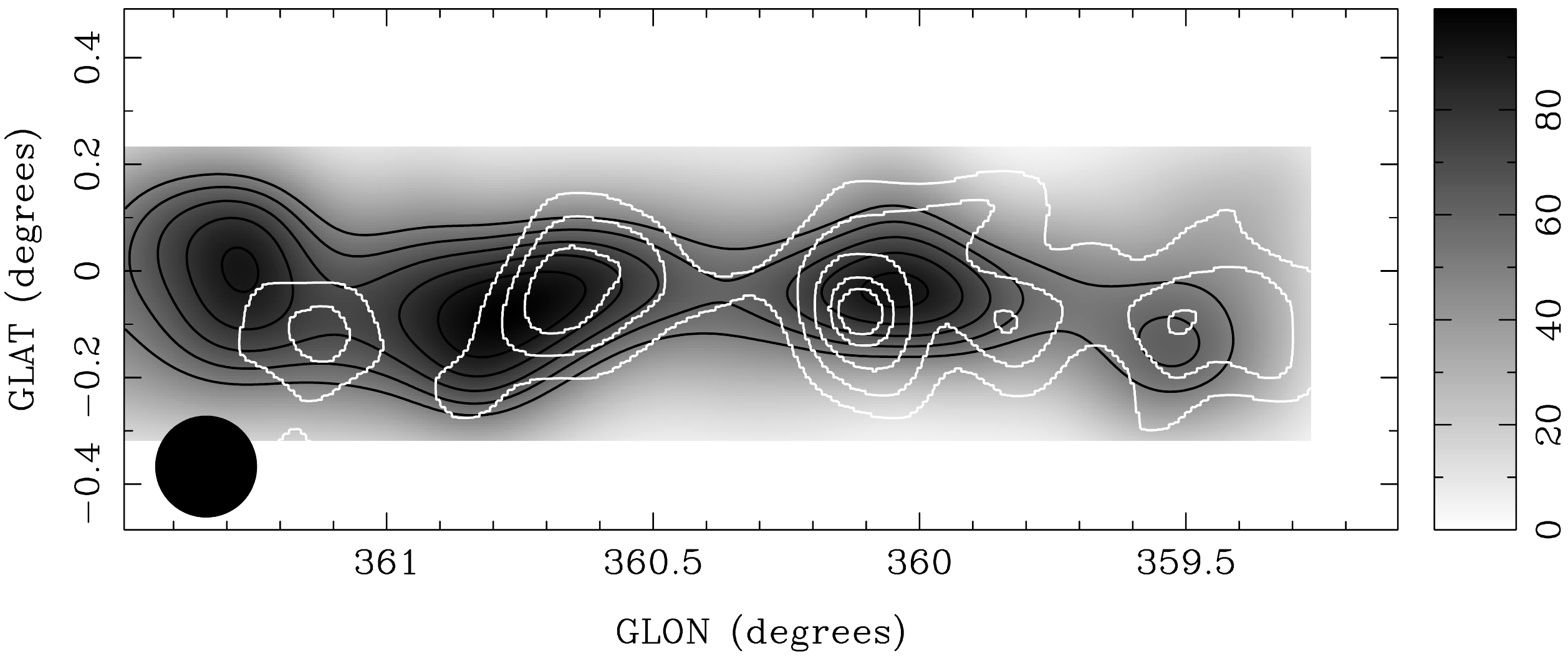,width=\columnwidth}
   \caption{Integrated emission image (over the range -200 to 200 km s$^{-1}$ in units of K km s$^{-1}$) of the HCN(1--0) emission line. Overlaid on the total intensity are (white) HESS contours at 290, 300, 320, 340 and 360 excess counts and (black) HCN(1--0) contours at 20\% to 90\% in steps of 10\%. The $11.3'$ beam, which the line emission has been convolved with is shown in the lower left-hand corner.}
   \label{fig:CMZ}
\end{figure}

Figure \ref{fig:CMZ}	 shows the result of integrating the HCN(1--0) emission from  -200 to 200 km s$^{-1}$ over the CMZ and smearing to the HESS resolution. The HCN(1--0) emission line is the brightest line observed in the CMZ dataset and Figure \ref{fig:CMZ} shows that, unlike the NH$_3$ (2,2) emission shown in Figure \ref{fig:HOPS}, there are several bright peaks (although the Sgr B GMC is certainly the brightest in Figure \ref{fig:CMZ}), corresponding to Sgr~C, Sgr~A, Sgr~B and the $l=1.3^\circ$ complex in order of increasing Galactic longitude.

Figure \ref{fig:CMZ} shows that the HCN(1--0) emission, broadly, follows that of the TeV gamma-rays in the GC region.  However, it does depart from the gamma-ray distribution in a couple of ways.  Firstly, the HCN(1--0) distribution is highly asymmetric about the GC (which is, indeed, characteristic of the CMZ region; \cite{Tsuboi1999}), more-so than even the TeV gamma-ray distribution. Additionally, the HCN(1--0) emission seems to follow a narrower distribution, in Galactic latitude than the gamma-rays -- although this is probably a function of the temperature and density (viz. $n_H\gtrsim10^4$~cm$^{-3}$ and $T_{ex}\sim70$~K) that the HCN(1--0) molecule traces.  This means that the HCN(1--0) molecules are tracing the dense clumps or cores of clouds, and not the broad distribution that lines such as the $^{12}$CO(1--0) emission line traces\footnote{It is well known that the $^{12}$CO(1--0) transition traces low density ($\sim100$~cm$^{-3}$) regions, which are typical for a large filling factor of the CMZ volume.}.

\section{Optical depth estimation}\label{sec:tau}
\begin{figure}[h]
   \centering
\epsfig{file=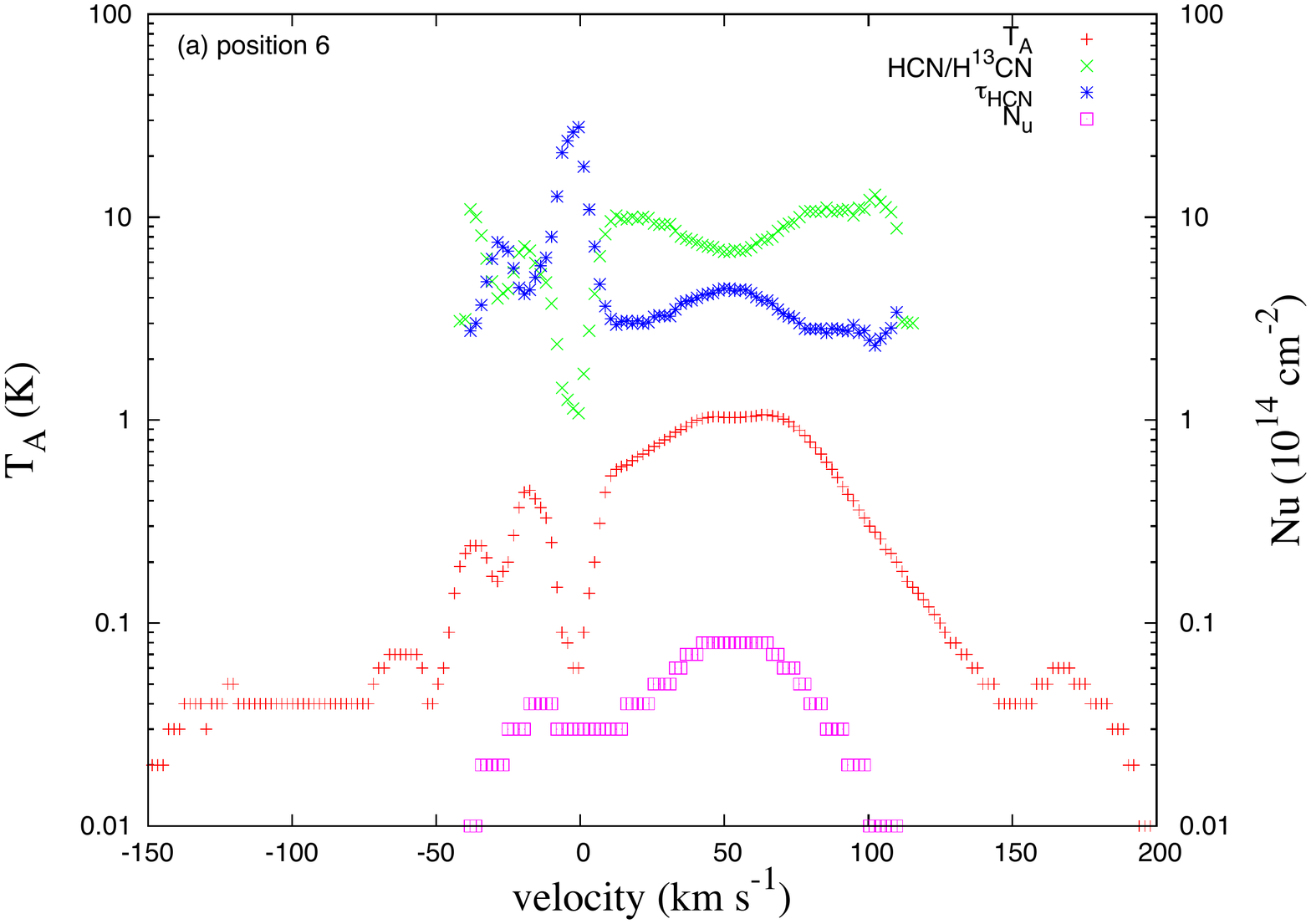,width=0.5\columnwidth}\epsfig{file=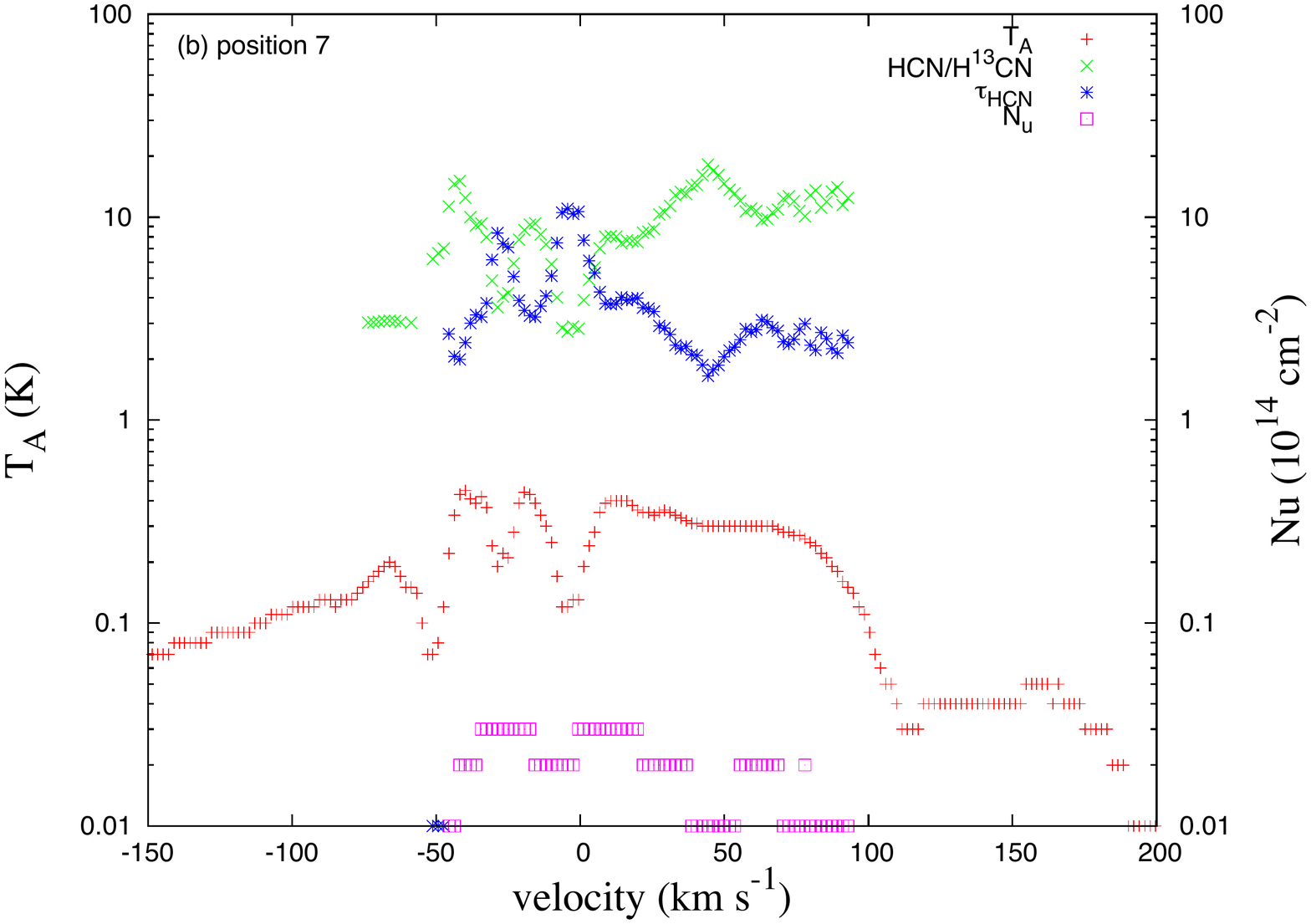,width=0.5\columnwidth}
   \caption{Plots of the antenna temperature ($T_A$), the ratio of isotopologue (HCN/H$^{13}$CN), the optical depth ($\tau$) of the HCN and the upper state column density ($N_u$) at (a) position 6 -- corresponding to Sgr~A -- and (b) position 7 -- corresponding to a region relatively free of gamma-ray and molecular line emission.  All data have been generated by convolution with a Gaussian beam equal to that of the HESS data (viz. $11.3'$) and $T_A$ and $N_u$ values can be read from the left and right y-axes respectively.}
   \label{fig:tau}
\end{figure}

From the CMZ survey, we have been able obtain images of the CMZ with several isotopologues so that, amongst other things, the estimation of the optical depth is possible. Figure \ref{fig:tau} (a) and (b) show various observational parameters that we have derived from the HCN/H$^{13}$CN(1--0) isotopologue pair. We have done this for nine regions within the CMZ, of which Figure \ref{fig:tau}(a) and (b) represent two, located at ($l,b$)=(0,0) -- towards Sgr~A, and ($l,b$)=(0,-0.6) -- containing little gamma-ray emission and labelled as position 6 and 7 in the figure respectively. In order to obtain the parameters on the same size scales that the TeV gamma-ray data samples, we have smeared the data to the resolution of the HESS data (FWHM of $11.3'$), clipped emission below the $5-\sigma$ noise level (apart from when determining the antenna temperature, $T^*_A$) and obtained the brightness temperature over the entire velocity range.  From this, we have derived several observational parameters, such as the profiles for HCN, H$^{13}$CN, their ratio, the optical depth ($\tau$) for HCN(1--0) derived from the HCN/H$^{13}$CN ratio, and the upper state column density ($N_u$, with no extinction correction).

Figure \ref{fig:tau}(a) and (b) shows that at these positions, the HCN line emission possesses an optical depth of about 2--4 at all velocities.  This is (i) found for all tracers in the CMZ sample; and (ii)  the same as that found by \cite{Tsuboi1999} using the CS (1--0) tracer which, in turn, will lead to an underestimation of the mass within the CMZ.  Additionally, this could possibly explain the offset of the peaks of gamma-ray emission in the GC and the better match with the peak emission of the NH$_3$ (2,2) and 8.5~GHz GBT emission: the gamma-ray emission is tracing the H{\sc ii} regions, but the tracers are less bright at these peaks due to the high optical depth.

The most powerful feature of this analysis, however, is that we will be able to (i) use the different isotopologue pairs (HCN/H$^{13}$CN, HCO$^+$/H$^{13}$CO$^+$, HNC/HN$^{13}$C and CS/$^{13}$CS) to explore and produce maps of the optical depth and other parameters which cover a range of excitation temperatures, number (and column) densities and chemistries.  This will enable us to more fully explore how well the TeV gamma-rays correlates with the molecules in the GC region.

\section{Conclusions}
We have compared the morphology of each molecular-line and radio continuum bands to the TeV gamma-ray emission in the CMZ at the HESS resolution. We have found that there is optically thick molecular gas present for all lines at all positions at all velocities within the CMZ, even where strong molecular-line and gamma-ray emission is not present. This will manifest itself as an under-estimation of the mass in the CMZ and may be responsible for the discrepancy between the peaks in the gamma-ray emission compared to the molecular-line emission. The fitting of telescopes, such as the Mopra telescope, with wide-band correlators has provided a boon to molecular line astronomy.  The results of surveys of the type that have been described and used here, can be especially powerful in gamma-ray astronomy's quest to elucidate the source of acceleration of protons and/or electrons to very high energies.

\end{document}